\documentclass[prl,nofootinbib,superscriptaddress,twocolumn]{revtex4}
\def\mysection#1{{\bf #1.} }
\def\mysections#1{{\bf #1.} }

\usepackage{amssymb}
\usepackage{amsmath}
\usepackage{graphicx}
\usepackage{longtable}
\usepackage{verbatim}
\usepackage{amsfonts}
\usepackage[bookmarksdepth=2]{hyperref}

\newcommand{\be}{\begin{eqnarray}}
\newcommand{\ee}{\end{eqnarray}}
\newcommand{\bea}{\begin{eqnarray}}
\newcommand{\eea}{\end{eqnarray}}
\newcommand{\beq}{\begin{eqnarray}}
\newcommand{\eeq}{\end{eqnarray}}
\def\beqa{\begin{eqnarray}}
\def\eeqa{\end{eqnarray}}
\newcommand{\no}{\nonumber}
\def\lsim{\mathrel{\rlap{\lower4pt\hbox{\hskip1pt$\sim$}}
    \raise1pt\hbox{$<$}}}         
\def\gsim{\mathrel{\rlap{\lower4pt\hbox{\hskip1pt$\sim$}}
    \raise1pt\hbox{$>$}}}         

\begin{document}

\vspace*{-30mm}

\title{\boldmath 2 Higgs or not 2 Higgs}

\author{Kfir Blum}\email{kblum@ias.edu}
\affiliation{Institute for Advanced Study, Princeton 08540, USA}

\author{Raffaele Tito D'Agnolo}\email{raffaele.dagnolo@sns.it}
\affiliation{Institute for Advanced Study, Princeton 08540, USA}
\affiliation{Scuola Normale Superiore and INFN, Piazza dei Cavalieri 7, 56126, Pisa, Italy}

\vspace*{1cm}

\begin{abstract}
\noindent
Motivated by recent results from the LHC experiments, we analyze Higgs couplings in two Higgs doublet models with an approximate $PQ$ symmetry. Models of this kind can naturally accommodate sizable modifications to Higgs decay patterns while leaving production at hadron colliders untouched. Near the decoupling limit, we integrate out the heavy doublet to obtain the effective couplings of the SM-like Higgs and express these couplings in a physically transparent way, keeping all orders in $(m_h/m_H)$ for small $PQ$ breaking. Considering supersymmetric models, we show that the effects on the Higgs couplings are considerably constrained.
\end{abstract}

\maketitle

\noindent
\mysection{Introduction}
\noindent
It is with a light heart that we assume, as a working hypothesis, that the recent measurements presented by CMS and ATLAS \cite{ATLAS, CMS} hint to an Higgs boson of mass $m_h\sim125$ GeV. 
Once the Higgs mass is obtained, the next guaranteed piece of information includes the Higgs production and decay rates in various modes. In the Standard Model (SM), given the Higgs mass, these quantities are completely determined theoretically and provide a probe of new physics. In this paper we take this promise for new information as motivation to study two Higgs doublet models (2HDMs). 2HDMs occur as part of the low energy particle content in many new physics scenarios. In such models, even in the presence of additional new particles, the tree level 2HDM-induced modifications to the SM-like Higgs couplings are irreducible and often make up the dominant contribution. 

Our study is useful in general for interpreting the pattern of deviations that can be expected in the context of 2HDMs. While we do not aim to explain in detail the current experimental state of affairs that is still inconclusive, before going into the main analysis we give a brief description of the situation at the time of writing. 
Presently, the data are consistent with $\mathcal{O}(1)$ enhancements with respect to a SM Higgs boson, for both gluon fusion (GF) and vector boson fusion (VBF) production channels, in the $\gamma \gamma$  \cite{ATLASphoton, CMSphoton}, and possibly also in the $ZZ$ and $WW$ decay modes~\cite{ATLASDiboson, CMSDiboson}. If the $\gamma \gamma$ rate (and potentially also $ZZ$ and $WW$ rates) turn out to be larger than in the SM, in both GF and VBF, then the multiple enhancements are more easily interpreted in terms of non-standard Higgs decay, rather than production. The simplest explanation being an $\mathcal{O}(1)$ reduction in the $hb\bar b$ coupling. 
However, a suppression in $hb\bar b$ must not be accompanied by a similar suppression in $ht\bar t$. Otherwise, without fortuitous interference between different new physics effects, GF production would be reduced by a similar amount. 
The situation is then such that (i) the couplings of $h$ to down-type quarks and to up-type quarks exhibit non-universal sensitivity to new physics, and (ii) the effect in the Higgs-bottom coupling is more pronounced than in the Higgs-top coupling. These requirements are met in 2HDMs with an approximate $PQ$ symmetry. 

In this class of models, some hierarchy, with one doublet parametrically heavier than the other, is motivated by experimental constraints. First, electroweak precision measurements limit the contribution of new physics to the $\rho$ parameter. This implies that new physics at the TeV scale should approximately conserve the diagonal $SU(2)_c$ custodial symmetry. 
Since $(H^-,A,H^+)$ transforms as a $\bf 3$ of $SU(2)_c$, the splitting between $m_{H^\pm}$ and $m_A$ is constrained in these models~\cite{Dine:2007xi,Mrazek:2011iu}. Furthermore the field $H\pm iA$ is charged under $PQ$, so the splitting between $m_{H^\pm}$ and $m_A$ is of the order of the $PQ$ breaking, that we are assuming is moderate. Second, the mass of the charged Higgs boson $H^\pm$ is constrained by its contribution to the decay $B\to X_s\gamma$ and recent calculations give the bound~\cite{Misiak:2006zs}
\beq m_{H^\pm}>295\,{\rm GeV}\eeq
at 95\%CL. Without accidental cancellations, this bound also applies to models with a richer particle spectrum, such as supersymmetry. As a result, it is natural to expect the whole doublet $(H^+,H+i A)$ to be parametrically heavier than $m_h\sim125$ GeV.

A mass hierarchy motivates an effective theory analysis of the 2HDM with only a single SM-like Higgs boson at the weak scale. In this paper we take on this analysis, adopting a somewhat different approach than in the previous literature. Focusing our attention to modified Higgs couplings, 
we present our results in a way that we find more transparently related to the symmetries of the theory from which the 2HDM and, eventually, the SM is assumed to descend. We show how these symmetries are reflected in observable Higgs signals and demonstrate the utility of our approach by easily deriving the modified Higgs couplings in several supersymmetric embeddings of the 2HDM.

%
\noindent
\mysection{2HDM analysis}
Consider a type-II 2HDM with $H_{1,2}\sim(1,2)_{+1/2}$, where only $H_1$ couples directly to $\bar Q_Ld_R$ and only $H_2$ couples directly to $\bar Q_Lu_R$ at high energies.  Neglecting leptons for now, the Lagrangian is~\cite{Gunion:2002zf,Branco:2011iw}
\beq\label{eq:V2hdm}-\mathcal{L}&=&H_1^\dag\mathcal{D}^2H_1+H_2^\dag\mathcal{D}^2H_2+m_1^2|H_1|^2+m_2^2|H_2|^2\no\\
&+&\frac{\lambda_1}{2}|H_1|^4+\frac{\lambda_2}{2}|H_2|^4+\lambda_3|H_1|^2|H_2|^2+\lambda_4|H_1\sigma_2 H_2|^2\no\\
&+&\Big\{\frac{\lambda_5}{2}(H_1^\dag H_2)^2+(H_1^\dag H_2)\left(m_{12}^2+\lambda_6|H_1|^2+\lambda_7|H_2|^2\right)\no\\
&+&Y_uH_2\epsilon \bar u_RQ_L+Y_dH_1^\dag \bar d_RQ_L+cc\Big\}\,.\eeq
The parameters $m_{12}^2,\,\lambda_{6},\,\lambda_{7}$ and $\lambda_5$ violate a $U(1)_{PQ}$ under which $(H_1^\dag H_2)$ has charge +1. A discrete $Z_2$ subgroup of this $U(1)_{PQ}$ controls the mixing between the two doublets. In this paper, we loosely refer to approximate $Z_2$ as the $PQ$ limit. Since the coupling $\lambda_5$ is even under the $Z_2$, it does not need to be small for our analysis to apply and indeed we will treat it collectively with other $Z_2$-even couplings. 
We parameterize spontaneous symmetry breaking (SSB) in a unitary gauge with
\beq H_1=\left(\begin{array}{c}h^+\\\frac{h_1+ia}{\sqrt{2}}\end{array}\right)\,,\;\;\;H_2=\left(\begin{array}{c}0\\\frac{h_2}{\sqrt{2}}\end{array}\right)\,,\;\;\;\langle h_2\rangle=v_2\,,
\eeq
where $a,h_1,h_2$ and the VEV $v_2$ are real. 

It is possible to diagonalize the Higgs mass matrix and express the couplings in terms of the rotation angle $\alpha$ connecting the interaction to the mass basis and of the ratio $\tan\beta$ between the VEVs of $H_2$ and $H_1$. This procedure gives $r_d\equiv \frac{vg_{hd\bar d}}{m_d}=-(\sin\alpha/\cos\beta)$, $r_u\equiv \frac{vg_{hu\bar u}}{m_d}=(\cos\alpha/\sin\beta)$ and $r_V\equiv \frac{vg_{hVV}}{2m_V^2}=\sin(\beta-\alpha)$. The trigonometric expressions for the $r_X$'s are useful as they provide the exact result and make apparent simple algebraic relations between them~\cite{Ferreira:2011aa}. They are less useful, however, if one looks for more insight into the underlying theory. 
Here, much in the spirit of~\cite{Mantry:2007ar,Randall:2007as}, we abandon the exact but somewhat less revealing $\alpha-\beta$ formulation in favor of a perturbative expansion, keeping track of the couplings in Eq.~(\ref{eq:V2hdm}) as we work out the solution. 

Our scheme is useful if the doublet $H_1$ is heavier than $H_2$, so that around the scale $m_h$ only $H_2$ is accessible. 
With this framework in mind we will obtain an effective action for $h_2$ to order $(B/M_1^2)^3$, where
\beq M_1^2=m_1^2+\frac{\lambda_{35}h_2^2}{2}\,,\;\;\;B=m_{12}^2+\frac{\lambda_7h_2^2}{2}\eeq
with\footnote{Compared with the basis of~\cite{Gunion:2002zf}, $(B/M_1^2)\sim1/\tan\beta$ and our $\lambda_{35}$ equals their $\lambda_{345}$. We 
regard insertions of $\lambda_7v^2,\;\lambda_6v^2$ on equal footing as insertions of $B$, as they carry the same PQ charge.} $\lambda_{35}=\lambda_3+\lambda_5$. 
We will not need to assume that $\lambda_{35}v^2\ll m_1^2$. This will improve the accuracy of our results for a mild hierarchy $m_h\lsim m_H$. 

Before proceeding to integrate out the heavy fields in $H_1$, we note some simplifying properties of the Lagrangian.
First, we assume that CP is conserved to a good approximation, and take all the potential couplings to be real. Under this assumption, scalars and pseudo scalars do not mix and we need only consider diagrams involving the two neutral scalars $h_1$ and $h_2$. Second, as defined in Eq.~(\ref{eq:V2hdm}), $\lambda_4$ projects neutral onto charged states and vice versa. It does not enter in tree diagrams with no charged external Higgs fields and we can ignore it in what follows. 
Third, working to $\mathcal{O}(B^3/M_1^6)$, we can ignore $\lambda_6$ and $\lambda_1$ that affect the results beginning at $\mathcal{O}(B^3/M_1^6)$ and $\mathcal{O}(B^4/M_1^8)$, respectively. 

Integrating out $h_1$ we obtain
\beq\label{eq:leff3}\!\!\!\!\!\!\!\!\!\!\!
-\mathcal{L}_{eff}
&=&\frac{1}{2}h_2\mathcal{D}^2h_2+\frac{1}{2}m_2^2h_2^2+\frac{\lambda_2}{8}h_2^4+\frac{Y_u}{\sqrt{2}}h_2t\bar t\no\\
&-&\frac{1}{2}Bh_2\frac{1}{\mathcal{D}^2+M_1^2}Bh_2-\frac{Y_b}{\sqrt{2}}b\bar b\frac{1}{\mathcal{D}^2+M_1^2}Bh_2.
\eeq
The interactions of the canonically normalized SM-like Higgs $h$ with the fermions and gauge bosons can be read off from (\ref{eq:leff3}), after accounting for wave function renormalization at $\mathcal{O}(B^2/M_1^4)$. In particular, the bottom-Higgs Lagrangian is given by
\beq\label{eq:leff4}
\frac{Y_b}{\sqrt{2}}b\bar b\frac{1}{\Box+M_1^2}B\left(v_2+\left(1-\frac{f'^2}{2}\right)h\right)
\eeq
with %
\beq v_2&=&v\left(1-\frac{f^2}{2v^2}\right),\;\;v^2=\frac{1}{\sqrt{2}G_F}\cong(246\,{\rm GeV})^2,\no\\
 f&=&\left\langle\frac{Bh_2}{M_1^2}\right\rangle,\;\;
f'=\frac{\partial f}{\partial v_2}.
\eeq
Using (\ref{eq:leff3}) and (\ref{eq:leff4}) we obtain:
\beq\label{eq:rfinal} r_b&=&\frac{vg_{hb\bar b}}{m_b}=\frac{1}{1-\frac{m_h^2}{M_1^2}}\left(1+\frac{\lambda_7v_2^2}{B}-\frac{\lambda_{35}v_2^2}{M_1^2}\right)\,,\no\\
r_t&=&\frac{vg_{ht\bar t}}{m_t}=1+\frac{B^2}{2M_1^4}\left(1-r_b^2\right)\,,\no\\
r_V&=&\frac{vg_{hVV}}{2m_V^2}=1-\frac{B^2}{2M_1^4}\left(1-r_b\right)^2\,.\eeq

The appearance of  terms $(m_h^2/M_1^2)$ in $r_b$ is due to the derivative operator in the effective vertex~(\ref{eq:leff4}). The $\Box$ operator is replaced by $\Box\to-m_h^2$ when acting on an external Higgs particle and by $\Box\to0$ when acting on the vacuum\footnote{We thank Nima Arkani-Hamed for a discussion on this point. This observation would also apply in a general effective theory analysis, like e.g. the one in~\cite{Mantry:2007sj}. }. Since $m_h$ corresponds to the physical mass, the $\Box$ operator automatically includes radiative corrections to $m_h$.

The deviation of $r_t$ from unity is parametrically small, beginning at $\mathcal{O}(B^2/M_1^4)$. The deviation of $r_V$ scales similarly. In contrast, the deviation of $r_b$ does not scale with $(B/M_1^2)$. It can be parametrically $\mathcal{O}(1)$ provided that either (i) $\lambda_7v^2\sim m_{12}^2$ or (ii) $\lambda_{35}v^2\sim m_1^2$. 

The condition $\lambda_7v^2\sim m_{12}^2$ implies that the hard and soft  breakings of the $PQ$ are comparable at the scale of SSB. Note that it is perfectly possible to have $\lambda_7v^2\sim m_{12}^2$ and $m_{12}^2\ll m_1^2$. For instance, if the theory at some high scale has $m_{12}^2\sim0$ but finite $\lambda_7$, we can expect $m_{12}^2\sim\lambda_7m_1^2/(4\pi)^2$ at scales below $m_1$. In this case we can have an $\mathcal{O}(1)$ correction to $r_b$ while the heavier doublet can be very heavy, $m_H\sim$ TeV. This shows that $r_b$ is a sensitive probe for hard breaking of the $PQ$~\cite{Randall:2007as}.

The second condition, $\lambda_{35}v^2\sim m_1^2$, implies that a sizable part of the mass of $H_1$ is driven by SSB. This is the relevant condition for models in which hard breaking of the $PQ$ is absent or small (like e.g. the MSSM). In this case, a discernible deviation of $r_b$ from unity implies a light second doublet with $m_H\sim v$. The corrections $\sim (m_h^2/M_1^2)$ coming from the derivative expansion can then be relevant; note that these terms correct $r_b$ with a definite positive sign. In the interesting case where soft $PQ$ breaking is also small, $B\ll m_h^2$, we can expand $r_b$ in $(B/M_1^2)$. In that case, we can replace $M_1^2\to m_H^2$, valid to $\mathcal{O}(B^2/M_1^4)$, in Eq.~(\ref{eq:rfinal}), obtaining 
\beq\label{eq:rbapp1} r_b\approx \left(1-\frac{m_h^2}{m_H^2}\right)^{-1}\left(1-\frac{\lambda_{35}v^2}{m_H^2}\right).\eeq
Eq.~(\ref{eq:rbapp1}) is correct to all orders in $(m_h^2/m_H^2)$ and $(\lambda_{35}v^2/m_H^2)$ and to $\mathcal{O}(B^2/M_1^4)$.

\begin{figure}[!t]
\begin{center}
\includegraphics[width=0.5\textwidth]{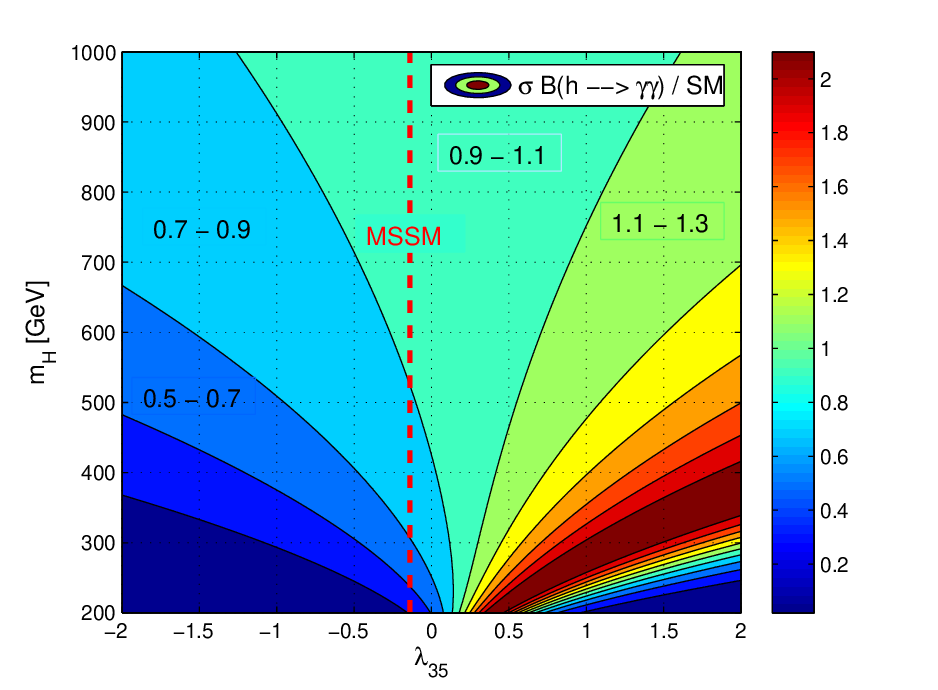}\end{center}
\caption{Contours of $\sigma\times BR(h\to\gamma\gamma)/({\rm SM})$ vs. $\lambda_{35}$ and $m_H$, for $m_h=125$ GeV and $\lambda_7=0$. (Recall $\lambda_{35}=\lambda_3+\lambda_5$.) The MSSM prediction, neglecting  loop corrections to the bottom Yukawa (but effectively including corrections to the Higgs potential), is shown by the dashed line.}
\label{fig:rb}
\end{figure}

Modifying the Higgs decay to bottom quarks affects the other search channels by changing the total width. A $125$ GeV Higgs in the SM has $BR(h\to b\bar b)\approx56\%$ so, for instance, the diphoton signal will be 
\beq\label{eq:dip}\frac{\sigma\times BR(h\to\gamma\gamma)}{\sigma\times BR(h\to\gamma\gamma)_{\rm SM}}\cong\frac{1}{1+0.56\left(r_b^2-1\right)}\,.\eeq
The effect on the $ZZ,WW$ final states is similar. 
In Fig.~\ref{fig:rb} we plot the diphoton enhancement from Eqs.~(\ref{eq:rbapp1})-(\ref{eq:dip}). The maximal enhancement is about a factor of two and is obtained for $m_H^2=m_h^2+\lambda_{35}v^2$. This means that, in the context of a 2HDM with an approximate $PQ$ and for order one couplings $\lambda_{35}\sim1$, taking the recent best fit ATLAS and CMS results~\cite{ATLASphoton, CMSphoton} at face value implies a light second doublet $m_H\sim300$ GeV. Note that in Eq.~(\ref{eq:dip}) we neglected the charged Higgs loop contribution to the coupling $h\gamma\gamma$. In the appendix we show that this contribution is indeed negligible for the range of $m_{H^\pm}$ that is consistent with $b\to X_s\gamma$.

Finally, using Eqs.~(\ref{eq:rfinal}) we give a quick prescription for computing the correction to $r_b$ in models with a type-II 2HDM at low energies. 
\begin{enumerate}
\item If the theory contains hard breaking of the $PQ$ via $\lambda_7$, then significant deviation is possible even for $m_H\sim$TeV in which case the leading effect is~\cite{Gunion:2002zf,Randall:2007as} 
\beq\label{eq:rbest} r_b\approx1+\frac{2}{1+2m_{12}^2/(\lambda_7v^2)}\,.\eeq
\item If there is little or no hard breaking of the $PQ$, $\lambda_7v^2\ll m_{12}^2$, then a modified $r_b$ requires a light second doublet. When soft $PQ$ breaking is also small, $B\ll m_h^2$, Eq.~(\ref{eq:rbapp1}) resums all powers of $(m_h^2/m_H^2)$ and $(\lambda_{35}v^2/m_H^2)$.
\end{enumerate}

So far we have neglected the Higgs coupling to leptons, but those can be added in a straightforward manner. If the doublet $H_1$ that couples to the down quarks couples also to the leptons, then $r_b=r_\tau$ and the change to the total width is amplified by a small factor $1+(m_\tau/m_b)^2/3\sim1.1$. 

\noindent
\mysection{Supersymmetric examples}
We now examine supersymmetric extensions of the SM with a 2HDM effective theory near the weak scale and extract the modifications to Higgs observables.

In supersymmetry, holomorphy of the superpotential requires a second Higgs doublet in order to couple the Higgs sector to both up- and down-type quarks. Identifying $H_d=i\sigma_2H_1^*,\,H_u=H_2$, the tree level quartic couplings of the MSSM are
\beq\lambda_1&=&\lambda_2=\frac{g^2+g'^2}{4},\;
\lambda_3=-\frac{g^2+g'^2}{4},\;
\lambda_4=\frac{g^2}{2},\no\\
\lambda_5&=&\lambda_6=\lambda_7=0\,.\eeq
The coupling $\lambda_{35}=\lambda_3+\lambda_5\approx-0.14$ is negative and so tends to increase $hb\bar b$. With $\lambda_{35}$ fixed and assuming $m_h=125$ GeV, Eq.~(\ref{eq:rbapp1}) tells us that, neglecting loop corrections to the bottom Yukawa, the value of $r_b$ depends only on $m_H$ with little sensitivity to the details of the supersymmetric spectrum. This result is in good agreement with FeynHiggs~\cite{Heinemeyer:1998yj}. The MSSM prediction for $r_b$ in this case translates to the vertical dashed line in Fig.~\ref{fig:rb}.

In some corners of the MSSM parameter space, our analysis ceases to give the leading result due to loop effects outside of the Higgs sector~\cite{Kane:1995ek,Gunion:2002zf,Carena:2000yi,Carena:2011aa}. The bottom Yukawa is corrected by stop-higgsino and sbottom-gaugino loops. The first contribution scales as $(\mu A_t/m^2_{soft})/(4\pi)^2$, and can dominate the coupling for $(\mu A_t/m^2_{soft})\sim$ few. The second contribution scales as $(Y_b\mu m_\lambda/m^2_{soft})/(4\pi)^2$ and can dominate for $(\mu m_\lambda/m^2_{soft})\gsim(4\pi)^2(B/m_H^2)\sim(4\pi)^2/\tan\beta$. 
In addition, light sfermions can affect the $h\gamma \gamma$ (and potentially $hGG$) vertex. In particular, very light staus could give an enhancment if $(M_1^2/B)$ (or $\tan\beta$) is so large that the tau Yukawa becomes $\mathcal{O}(1)$~\cite{Carena:2011aa}. This defines a lower limit for $(B/M_1^2)$ where we can neglect bottom and tau loop corrections: $(B/M_1^2)\gsim(\sqrt{2}m_b/v)\sim1/40$. 

It is interesting to ask whether simple extensions of the MSSM that accommodate a large Higgs mass in a more natural way can also reduce $hb\bar b$. We briefly examine three such models, the NMSSM, the BMSSM and a $U(1)_X$.
Considering a $Z_3$ version of the NMSSM, we write the superpotential
\beq\label{eq:nmssmW}\mathcal{W}&=&\lambda SH_uH_d+\frac{\kappa}{3} S^3\,. \eeq
We assume that $S$ is somewhat heavy so that at low energy the theory can be described by the 2HDM.
The coupling $\lambda_3$ is then given by
\beq\label{eq:nmssm}\lambda_3&=&-\frac{g^2+g'^2}{4}+|\lambda|^2\approx-0.14+0.5\left|\frac{\lambda}{0.7}\right|^2\, \eeq
and is larger than in the MSSM. Still, as can be seen from Fig.~\ref{fig:rb}, for this effect alone to achieve $r_b<1$, $\lambda>0.7$ is required, above the limit of perturbative unification~\cite{Hall:2011aa}. 

Adding to Eq.~(\ref{eq:nmssmW}) the supersymmetric mass terms $\mu H_uH_d$ and $\frac{\Lambda}{2}S^2$ with $\Lambda\gg\mu$ produces the BMSSM~\cite{Dine:2007xi}. The spurion $\epsilon_1=(\lambda^2\mu^*/\Lambda)$ carries $PQ$ charge -1 and induces $\lambda_6=\lambda_7=-2\epsilon_1$.  For $(B/M_1^2)\lsim1/10$, $\epsilon_1$ could decrease $r_b$ while making a negligible correction to $m_h$~\cite{Randall:2007as}. 

Next, consider a gauge $U(1)_X$ extension under which the Higgs fields are charged, $q_{H_u}=q_{H_d}\equiv q_H$. 
The scalar potential receives a correction
\beq V&=&V_{MSSM}+\frac{g_X^2q_H^2}{2}\left(|H_2|^2+|H_1|^2\right)^2\eeq
Leading to $\lambda_3\to\lambda_3+g_X^2q_H^2$. 
The modification to the $hb\bar b$ coupling in this example is in fact limited by the Higgs mass: since $\delta m_h^2\sim 2g_X^2q_H^2v^2$, we should impose $g_X^2q_H^2<m_h^2/(2v^2)\approx0.12$. We therefore do not expect large deviations from the MSSM tree level predictions.

\noindent
\mysection{Conclusions}
We have shown that a 2HDM close to the decoupling limit with an approximate $PQ$ symmetry (or a $Z_2$ subgroup of it) can produce $\mathcal{O}(1)$ deviations in the lightest Higgs couplings to SM particles. Integrating out the heavier Higgs, we presented the effective couplings of the light SM-like Higgs in a physically transparent way. Keeping derivative operators in the expansion allowed us to include automatically radiative corrections to the lightest Higgs mass, important in the limit of small $PQ$ breaking but a mild hierarchy $m_H\sim m_h$. 

Our results are applicable to any type-II 2HDM not far from decoupling. In particular, considering the MSSM and some of its extensions, our analysis elucidates why it is hard to enhance $h\to\gamma \gamma$. If the current experimental hints are confirmed as the statistics increases, these results will allow to set bounds on the heavy Higgs mass $m_H$ in large regions of these models parameter space.

Assuming a 2HDM with an approximate $PQ$ and natural couplings, taking the recent best fit ATLAS and CMS results~\cite{ATLASphoton, CMSphoton} at face value implies a not too-heavy second doublet $m_H\sim300$ GeV. Further support for this possibility should come from better measurements of the SM-like Higgs decay patterns, where the generic type-II 2HDM predicts similar enhancements in gluon fusion and vector boson fusion production for $\gamma\gamma,\,ZZ$ and $WW$ final states, with correspondingly decreased $h\to b\bar b$. 

\mysections{Acknowledgments}
We thank Janie Chan, David E. Kaplan, Rouven Essig, JiJi Fan and Neal Weiner for useful discussions. We are grateful to Nima Arkani-Hamed for many insightful discussions and for comments on the manuscript. 
KB is supported by DOE grant DE-FG02-90ER40542.

\noindent
\mysection{Appendix: Charged Higgs contribution to $h\gamma\gamma$}
The decay $h\to \gamma \gamma$ is mediated by a dimension five Lagrangian, that we parametrize by~\cite{Manohar:2006gz}
\beq
\mathcal{L}_{\gamma}=-\frac{2\pi\alpha vc_\gamma }{\Lambda^2}hF_{\mu\nu}F^{\mu\nu}-\frac{2\pi\alpha v\tilde c_\gamma}{\Lambda^2}h\tilde F_{\mu\nu}F^{\mu\nu}\,. 
\eeq
In the absence of CP-violation, $\tilde c_\gamma=0$. 
The contribution of the charged Higgs loop is given by~\cite{Djouadi:2005gi}
\beq\label{eq:gamloop}\frac{c_\gamma}{\Lambda^2}&=&-\frac{\lambda_{hH^\pm H^\pm}}{v}\,\frac{f\left(\tau\right)}{24\pi^2m^2_{H^\pm}}\,,\;\;\;\tau=\frac{m_h^2}{4m^2_{H^\pm}}\eeq
with
\beq f(r)=3\frac{\arcsin^2(\sqrt{r})-r}{r^2}\cong1+0.6\,r,\eeq
where the last approximation is valid to one percent accuracy for $m_h^2/(4m^2_{H^\pm})<0.2$ or, in case of $m_h=125$ GeV, for $m_{H^\pm}>140$ GeV. The coupling $(\lambda_{hH^\pm H^\pm}/v)$ can be obtained from Eq.~(\ref{eq:V2hdm}),
\beq
\frac{\lambda_{hH^\pm H^\pm}}{v}&=&\lambda_{34}
\eeq
with $\lambda_{34}=\lambda_3+\lambda_4$ to $\mathcal{O}(B^2/M_1^4)$. In the MSSM, for example, $\lambda_{34}=(g^2-g'^2)/4\approx0.07$.

Adding this correction to the SM W and top loop contributions gives
\beq\!\!\!\!\frac{\Gamma(h\to\gamma\gamma)}{\Gamma(h\to\gamma\gamma)_{SM}}&=\left|\frac{r_V\mathcal{I}^W+r_t\frac{4}{3}\left(1-\frac{\alpha_s}{\pi}\right)\mathcal{I}^t}{\mathcal{I}^\gamma}-\frac{4\pi^2v^2c_{\gamma}}{\Lambda^2\mathcal{I}^\gamma}\right|^2,
\eeq
with the loop function $\mathcal{I}^\gamma=\mathcal{I}^W+\frac{4}{3}\left(1-\frac{\alpha_s}{\pi}\right)\mathcal{I}^t$ taken from~\cite{Manohar:2006gz}.
For $m_h=125$ GeV we find 
$\mathcal{I}^W=-2.09,\;\mathcal{I}_t=0.34,\;\mathcal{I}^\gamma=-1.645$. 
Unless the charged Higgs is very light, or the relevant quartic couplings are large, the contribution to the $h\gamma\gamma$ coupling makes a negligibly small correction to the SM terms:
\beq
\label{eq:GammaBR}&\frac{\Gamma(h\to\gamma\gamma)}{\Gamma(h\to\gamma\gamma)_{SM}}
\cong\no\\
&\left|1.27r_V-0.27r_t-0.05\left(\frac{\lambda_{hH^\pm H^\pm}}{v}\right)\left(\frac{m_{H^\pm}}{350\,\rm GeV}\right)^{-2}\right|^2.\no\eeq
%



\begin{thebibliography}{01}
\vspace*{3mm}

\bibitem{ATLAS} 
ATLAS~Collaboration,
  arXiv:1202.1408 [hep-ex].

\bibitem{CMS} 
 CMS~Collaboration,
  arXiv:1202.1488 [hep-ex].
    
 \bibitem{ATLASphoton} 
  ATLAS~Collaboration,
  arXiv:1202.1414 [hep-ex].

\bibitem{CMSphoton}
  CMS Collaboration,
  arXiv:1202.1487 [hep-ex].

\bibitem{ATLASDiboson} 
  ATLAS~Collaboration,
  arXiv:1202.1415 [hep-ex].
G.~Aad {\it et al.}  [ATLAS Collaboration],
  arXiv:1112.2577 [hep-ex].
\bibitem{CMSDiboson}
CMS~Collaboration,
  arXiv:1202.1489 [hep-ex].
CMS~Collaboration,
  arXiv:1202.1416 [hep-ex].

\bibitem{Dine:2007xi} 
  M.~Dine, N.~Seiberg and S.~Thomas,
  Phys.\ Rev.\ D {\bf 76}, 095004 (2007)
  [arXiv:0707.0005 [hep-ph]].
  
\bibitem{Mrazek:2011iu} 
  J.~Mrazek, A.~Pomarol, R.~Rattazzi, M.~Redi, J.~Serra and A.~Wulzer,
  Nucl.\ Phys.\ B {\bf 853}, 1 (2011)
  [arXiv:1105.5403 [hep-ph]].

\bibitem{Misiak:2006zs} 
  M.~Misiak, H.~M.~Asatrian, K.~Bieri, M.~Czakon, A.~Czarnecki, T.~Ewerth, A.~Ferroglia and P.~Gambino {\it et al.},
  Phys.\ Rev.\ Lett.\  {\bf 98}, 022002 (2007)
  [hep-ph/0609232].
  
\bibitem{Gunion:2002zf}
  J.~F.~Gunion and H.~E.~Haber,
  Phys.\ Rev.\ D {\bf 67} (2003) 075019
  [hep-ph/0207010].
  
\bibitem{Branco:2011iw} 
  G.~C.~Branco, P.~M.~Ferreira, L.~Lavoura, M.~N.~Rebelo, M.~Sher and J.~P.~Silva,
  arXiv:1106.0034 [hep-ph].
  
\bibitem{Ferreira:2011aa} 
  P.~M.~Ferreira, R.~Santos, M.~Sher and J.~P.~Silva,
  arXiv:1112.3277 [hep-ph].
  
\bibitem{Mantry:2007ar} 
  S.~Mantry, M.~Trott and M.~B.~Wise,
  Phys.\ Rev.\ D {\bf 77}, 013006 (2008)
  [arXiv:0709.1505 [hep-ph]].

\bibitem{Randall:2007as} 
  L.~Randall,
  JHEP {\bf 0802}, 084 (2008)
  [arXiv:0711.4360 [hep-ph]].
  
\bibitem{Mantry:2007sj} 
  S.~Mantry, M.~J.~Ramsey-Musolf and M.~Trott,
  Phys.\ Lett.\ B {\bf 660}, 54 (2008)
  [arXiv:0707.3152 [hep-ph]].
  
 \bibitem{Heinemeyer:1998yj} 
  S.~Heinemeyer, W.~Hollik and G.~Weiglein,
  Comput.\ Phys.\ Commun.\  {\bf 124}, 76 (2000)
  [hep-ph/9812320];
  S.~Heinemeyer, W.~Hollik and G.~Weiglein,
  Eur.\ Phys.\ J.\ C {\bf 9}, 343 (1999)
  [hep-ph/9812472];
  G.~Degrassi, S.~Heinemeyer, W.~Hollik, P.~Slavich and G.~Weiglein,
  Eur.\ Phys.\ J.\ C {\bf 28}, 133 (2003)
  [hep-ph/0212020];
  M.~Frank, T.~Hahn, S.~Heinemeyer, W.~Hollik, H.~Rzehak and G.~Weiglein,
  JHEP {\bf 0702}, 047 (2007)
  [hep-ph/0611326].

\bibitem{Carena:2011aa} 
  M.~Carena, S.~Gori, N.~R.~Shah and C.~E.~M.~Wagner,
  arXiv:1112.3336 [hep-ph].
 
\bibitem{Kane:1995ek} 
  G.~L.~Kane, G.~D.~Kribs, S.~P.~Martin and J.~D.~Wells,
  Phys.\ Rev.\ D {\bf 53}, 213 (1996)
  [hep-ph/9508265].

\bibitem{Carena:2000yi} 
  M.~S.~Carena, J.~R.~Ellis, A.~Pilaftsis and C.~E.~M.~Wagner,
  Nucl.\ Phys.\ B {\bf 586}, 92 (2000)
  [hep-ph/0003180].
 
\bibitem{Hall:2011aa} 
  L.~J.~Hall, D.~Pinner and J.~T.~Ruderman,
  arXiv:1112.2703 [hep-ph].

\bibitem{Manohar:2006gz} 
  A.~V.~Manohar and M.~B.~Wise,
  Phys.\ Lett.\ B {\bf 636}, 107 (2006)
  [hep-ph/0601212].
 
 \bibitem{Djouadi:2005gi} 
  A.~Djouadi,
  Phys.\ Rept.\  {\bf 457}, 1 (2008)
  [hep-ph/0503172];
  A.~Djouadi,
  Phys.\ Rept.\  {\bf 459}, 1 (2008)
  [hep-ph/0503173].
  

%
%
%
%
%
%
%
%
%
%
%
%
%
%
%
  
 \end{thebibliography}
\end{document}